\documentclass[10pt,
			   aps,
			   prl,
			   twocolumn,
			   nofootinbib,
			   showkeys,           % show keywords
			   superscriptaddress  % for auto numbered affils
			   ]{revtex4-1}

\usepackage{amssymb,amsmath}
\usepackage{cmap}
\usepackage{graphicx}
\usepackage{color}
\usepackage{hyperref}
\usepackage{listings}
\usepackage{booktabs}
\usepackage{multirow}
\hypersetup{
    unicode=false,           % non-Latin characters in Acrobat’s bookmarks
    pdftoolbar=true,         % show Acrobat’s toolbar?
    pdfmenubar=true,         % show Acrobat’s menu?
    pdffitwindow=false,      % window fit to page when opened
    pdfstartview={FitH},     % fits the width of the page to the window
    pdfauthor={Pavel Belov}, % author
    colorlinks=true,         % false: boxed links; true: colored links
    linkcolor=blue,          % color of internal links
    citecolor=blue,           % color of links to bibliography
}
\usepackage{dsfont}
\usepackage{physics}

\usepackage[english]{babel}

\graphicspath{{fig/}}

\def\vec{\mathbf}

\begin{document}

\title{Longitudinal electromagnetic waves with extremely short wavelength}

\newcommand{\affilITMO}{Department of Physics and Engineering, ITMO University, Kronverksky pr. 49, 197101, Saint Petersburg, Russia}

\author{Denis Sakhno}
\affiliation{\affilITMO}
\author{Eugene Koreshin}
\affiliation{\affilITMO}
\author{Pavel A. Belov}
\email{belov@metalab.ifmo.ru}
\affiliation{\affilITMO}

\begin{abstract}
Electromagnetic waves in vacuum and most materials have transverse polarization. Longitudinal electromagnetic waves with electric field parallel to wave vector are very rare and appear under special conditions in a limited class of media, for example in plasma. 
In this work, we study the dispersion properties of an easy-to-manufacture metamaterial consisting of two three-dimensional cubic lattices of connected metallic wires inserted one into another, also known as an interlaced wire medium. It is shown that the metamaterial supports longitudinal waves at extremely wide frequency band from very low frequencies up to the Bragg resonances of the structure.
The waves feature unprecedentedly short wavelengths comparable to the period of the material.
The revealed effects highlight spatially dispersive response of interlaced wire medium and provide a route toward generating electromagnetic fields with strong spatial variation.
\end{abstract}

%\keywords{interlaced wire medium, metamaterial, spatial dispersion.}
%keywords are no longer used in Phys Rev papers (MG)

\maketitle

%\section{Introduction}
Elastic waves may have either longitudinal or transverse polarization: the acoustic (compression) waves are longitudinal while the sheer stress waves are usually transverse. The electromagnetic waves are in many aspects similar to the elastic ones. However, most of electromagnetic waves are transverse.
The longitudinal wave has wave vector $\vec{k}$ parallel to electric field $\vec{E}$.
Substitution of such conditions into Maxwell's equations for non-magnetic isotropic media immediately leads to $\vec H=0$ and $\varepsilon =0$.
This means that the longitudinal electromagnetic waves may exist in homogeneous media only if the dielectric permittivity is equal to zero.
Such case can be reached in plasmas or plasma-like media also known as epsilon-near-zero (ENZ) materials \cite{Liberal2017,ENZReview2,ENZReview1}.

The longitudinal wave in such case is called bulk plasmon \cite{Ginzburgbook} and it exists at the particular frequency $\omega_p$ called plasma frequency defined by the equation $\varepsilon (\omega_p)=0$.
Due to frequency dispersion, the condition $\varepsilon(\omega)=0$ can be satisfied only at a fixed frequencies. However, spatial dispersion effects (nonlocality) manifested as the dependence of permittivity on the wave vector of propagating wave $\varepsilon (\omega, \vec{k})$, allows bulk plasmons to exist within a certain very narrow frequency band~\cite{Ginzburgbook,PIAZZA1984}.

% The equation $$\varepsilon (\omega, \vec{k})=0$$ has solution $\vec {k} (\omega)=...$

In natural media the effects of spatial dispersion are extremely weak since the period of the crystal is significantly smaller than the wavelength.
The bandwidth of the bulk plasmon band is only a few percents of the respective frequency \cite{Ginzburgbook,PIAZZA1984}.
Metamaterials which are artificially synthesized media~\cite{WegenerNatureRP2019} feature typically stronger spatial dispersion effects since their periods are comparable to $10^{-1}-10^{-2} \lambda$. For instance, connected wire medium also known as artificial plasma \cite{Brown1} supports bulk plasmons within 7$\%$ band near the plasma frequency \cite{Pendry1,Pendry2,silveirinha2009plasma}.

Spatial dispersion effects are boosted when the ratio between the period of the structure and wavelength increases. For example, the effect of spatial-dispersion-induced birefringence quantified by the difference of the refractive indices for two orthogonal polarizations in a metamaterial with cubic symmetry can reach 0.13 in resonant metamaterials as compared to $10^{-5}$ in the natural crystals such as CuI or NaI~\cite{gorlach2016giant}. However, the discussed spatial dispersion effects are observed within quite narrow frequency range close to the characteristic resonances of a metamaterial.

The exception from this general rule is provided by the non-connected wire media \cite{simovski2012wire} which are metamaterials formed by parallel arrays of infinitely long wires disconnected from each other. The non-connected wire media feature very strong spatial dispersion within extremely wide frequency range including the long-wavelength limit \cite{WMSD}. Their nonlocal electromagnetic response results in the diffractionless transverse electromagnetic waves with arbitrary transverse wave-vectors which can be used for subwavelength imaging \cite{WMlens} and improvement of magnetic resonance imaging systems \cite{fish}.

\begin{figure}[h] %--------------------------------FIGURE!
	\begin{minipage}[h]{0.9\linewidth}
		\center{\includegraphics[width=1\textwidth]{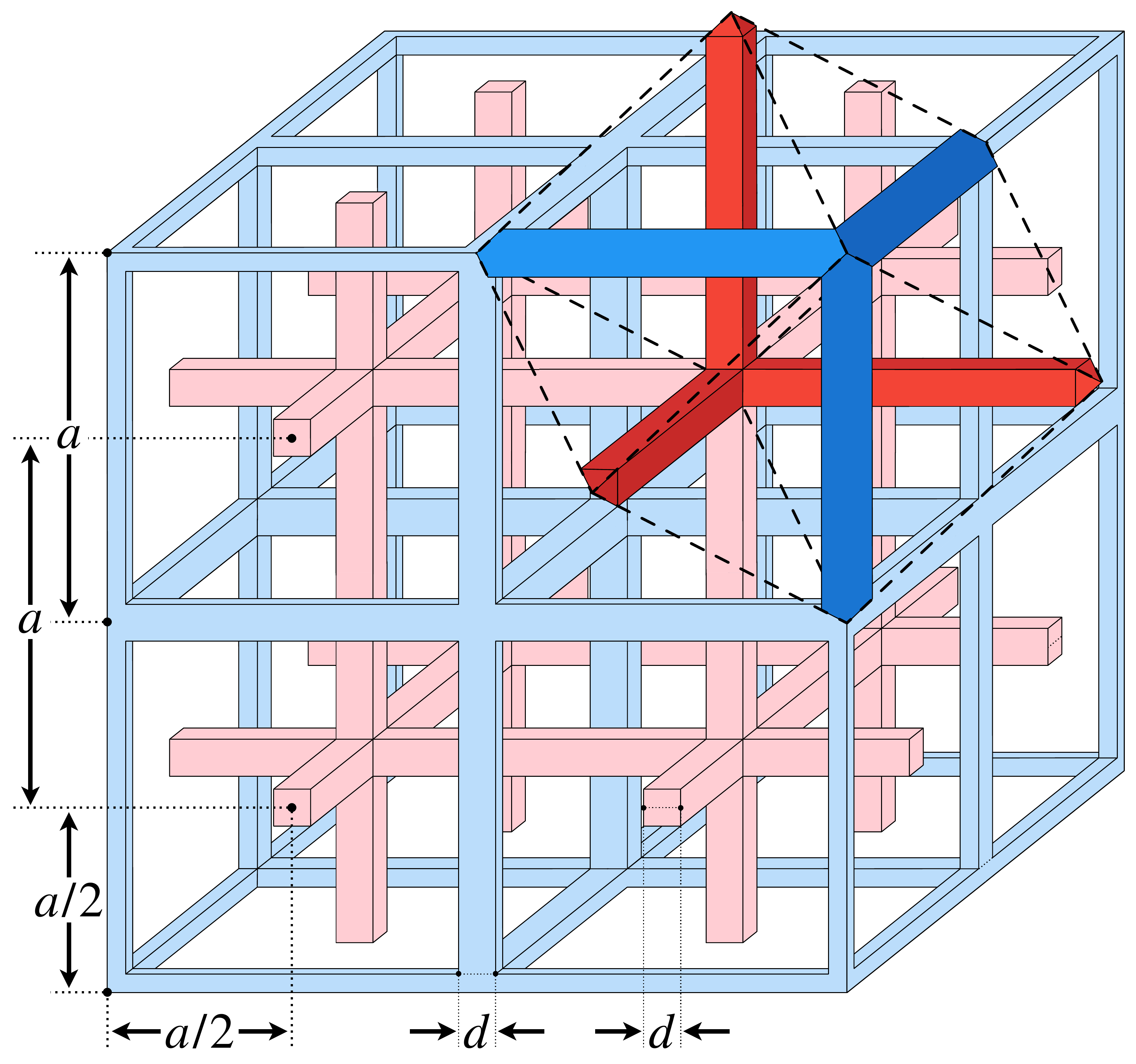} \\ (a)}
	\end{minipage}
	\vfill
	\begin{minipage}[h]{0.49\linewidth}
		\center{\includegraphics[width=1\textwidth]{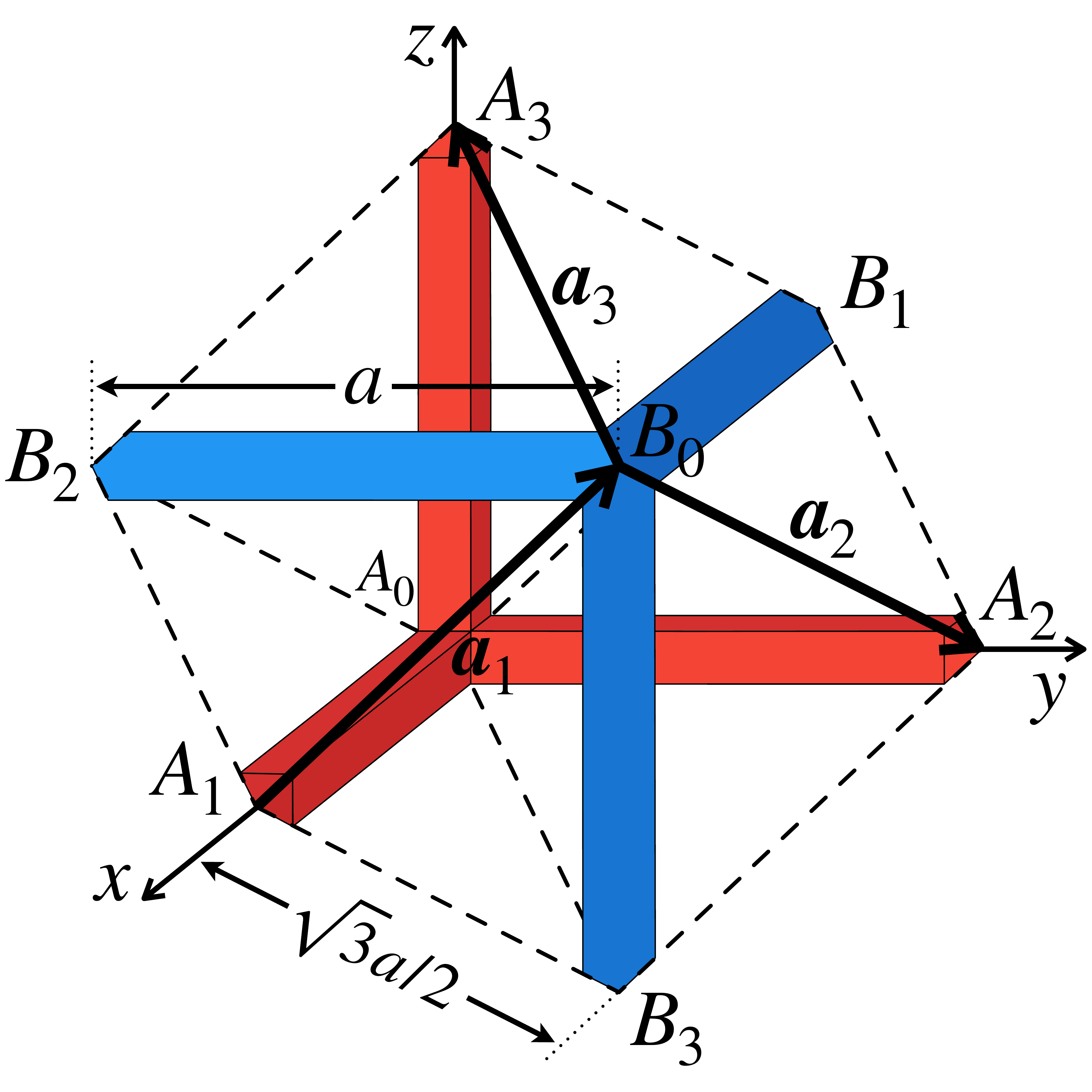} \\ (b)}
	\end{minipage}
	\hfill
	\begin{minipage}[h]{0.49\linewidth}
		\center{\includegraphics[width=1\textwidth]{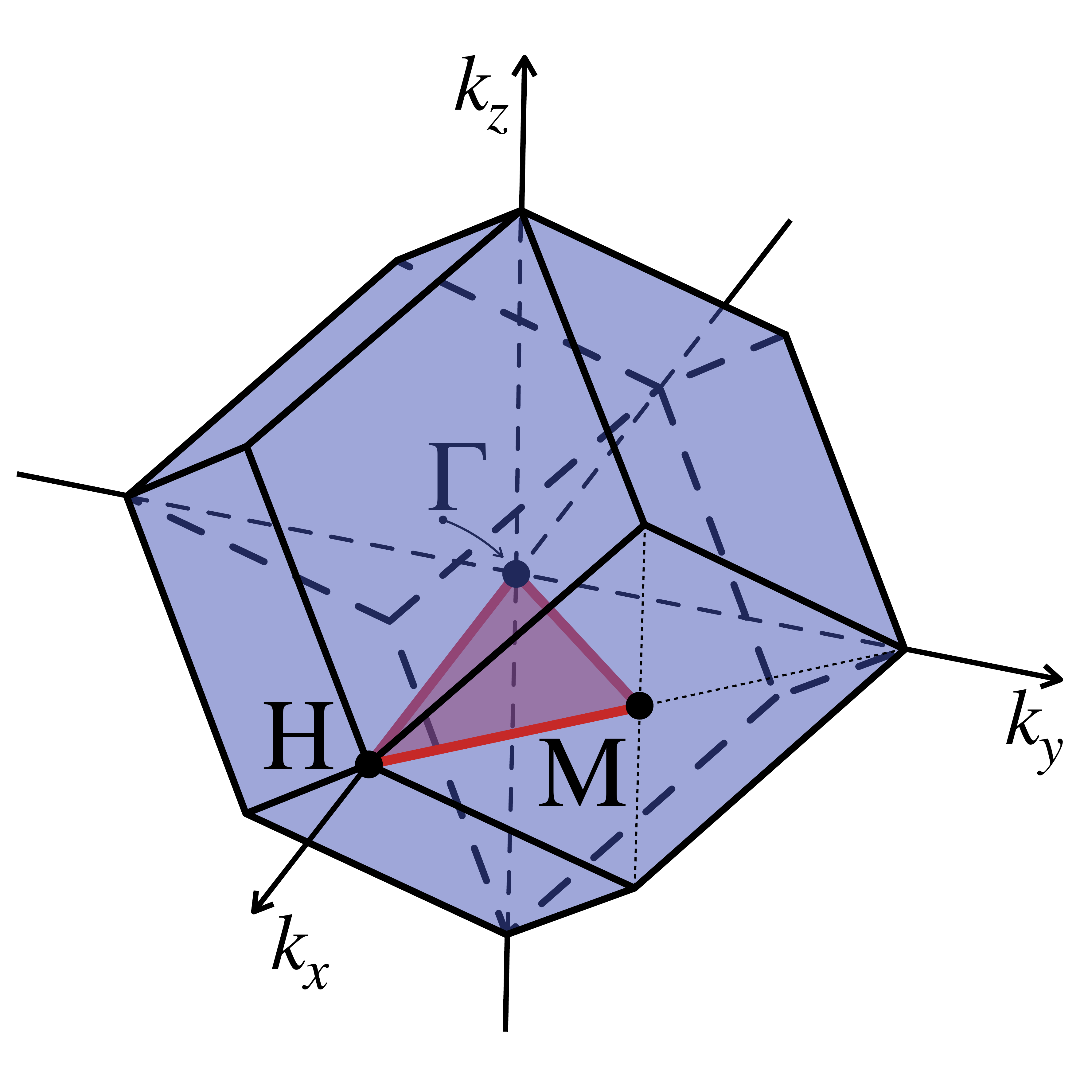} \\ (c)}
	\end{minipage}
	\caption{(a) Geometry of the interlaced wire medium. (b) Rhombohedral unit cell of the metamaterial. Translation vectors coordinates: $A_1B_0=\left(-a/2,a/2,a/2\right)^\mathrm{T}$, $B_0A_2=\left(-a/2,a/2,-a/2\right)^\mathrm{T}$, $B_0A_3=\left(-a/2,-a/2,a/2\right)^\mathrm{T}$. (c) Brillouin zone corresponding to the unit cell. Points coordinates: $\Gamma=\left(0,0,0\right)^\mathrm{T}$,  $\mathrm{H}=\left(2\pi/a,0,0\right)^\mathrm{T}$, $\mathrm{M}=\left(\pi/a,\pi/a,0\right)^\mathrm{T}$.}
	\label{fig:unitcell}
\end{figure}

In this Letter, we investigate the properties of a special class of wire media, so called interlaced wire media \cite{fan2007nonmax,silveirinha2017lighttun,Powell2021}. The geometry of the medium is shown in Fig. \hyperref[fig:unitcell]{\ref{fig:unitcell}(a)}:
it consists of two 3d wire meshes embedded one into another without an electrical connection between the meshes. Due to this, the interlaced wire medium at low frequencies in many ways differs from single 3d wire metamaterial \cite{Brown1,Pendry1,Pendry2,silveirinha2009plasma}. Basically, 3d wire metamaterial feature inductive behavior whereas interlaced wire medium following to \cite{fan2007nonmax} behaves more like a transmission line with both inductive and capacitive behavior, where the effective capacitance is associated with the two non-connected sub-meshes.

Below, we demonstrate that such structure supports longitudinal waves within extremely wide frequency band starting from very low frequencies up to at least frequency corresponding to the wavelength which is 6 times greater than the period of the metamaterial. Due to the pronounced spatially dispersive response of the interlaced wire medium, the equation $\varepsilon (\omega, \vec{k})=0$ has solutions nearly at all frequencies below the first Bragg resonance. Interestingly, the wave vectors of the waves are very large pointing to the corners of the first Brillouin zone of the metamaterial. It should be stressed that such behaviour is quite unusual for the electromagnetic waves since in the majority of materials low frequencies are related to short wave vectors of the wave such that isofrequency contours are centered around $\Gamma=(0,0,0)^T$ point of the Brillouin zone.

% \section{\label{sec:huge-k} Huge wave vectors}

We consider a metamaterial consisting of two identical perfectly conducting cubic wire meshes with a period $a$ and $d\times d$ square cross section ($d=0.1a$) inserted one into the other (Fig. \hyperref[fig:unitcell]{\ref{fig:unitcell}(a)}) and fixed in the position with maximal distance between the networks nodes (the networks displacement vector $A_0B_0=0.5a\left(1,1,1\right)^\mathrm{T}$, see Figs. \hyperref[fig:unitcell]{\ref{fig:unitcell}(a)} and \hyperref[fig:unitcell]{\ref{fig:unitcell}(b)}). 
The entire structure is placed in an isotropic host medium (vacuum). We consider here two identical connected wire meshes in contrary to the previous work~\cite{silveirinha2017lighttun} where wires in the meshes had different radii.

The structure has body-centered cubic (bcc) symmetry (in nature some metals like lithium (Li), chromium (Cr), tungsten (W) and some others have such crystal structure, canonical bcc) and belongs to $O_h$ symmetry group. Figure~\hyperref[fig:unitcell]{\ref{fig:unitcell}(b)} shows the geometry of the unit cell for the interlaced wire metamaterial and a way how it is related to the structure of the medium (Fig. \hyperref[fig:unitcell]{\ref{fig:unitcell}(a)}). This unit cell is a rhombohedral, i.e. a hexagon with equal rhombuses at all faces, all ribs are equal to $\sqrt{3}a/2$. One of two diagonals at each face contains a metal wire of length $a$. This rhombohedral cell is well known in solid state physics as \textit{primitive} cell for bcc crystal structures.
%(the Wigner-Seitz cell can also be chosen). The primitive cell was taken for the following application of a quasistatic theory. 
The coordinates of the vertices can be written as:
\begin{equation}
    \begin{array}{ll}
    A_0=\left(0,0,0\right)^\mathrm{T} &
    B_0=\left(a/2,a/2,a/2\right)^\mathrm{T}\\
    A_1=\left(a,0,0\right)^\mathrm{T} &
    B_1=\left(-a/2,a/2,a/2\right)^\mathrm{T}\\
    A_2=\left(0,a,0\right)^\mathrm{T} &
    B_2=\left(a/2,-a/2,a/2\right)^\mathrm{T}\\
    A_3=\left(0,0,a\right)^\mathrm{T} &
    B_3=\left(a/2,a/2,-a/2\right)^\mathrm{T}\\
    \end{array}
    \label{eq:vertex}
\end{equation}

The translation vectors:
\begin{equation}
    \begin{matrix}
    \boldsymbol{a}_1 = A_1B_0,&
	\boldsymbol{a}_2 = B_0A_2,&
	\boldsymbol{a}_3 = B_0A_3.
    \end{matrix}
    \label{eq:dir-vectors}
\end{equation}

The interlaced wire medium with the described unit cell has a complex twelve-sided Brillouin zone (rhombic dodecahedron) shown in Fig. \hyperref[fig:unitcell]{\ref{fig:unitcell}(c)}.

\begin{figure}[h!] %--------------------------------FIGURE!
	\begin{minipage}[h]{1\linewidth}
		\center{\includegraphics[width=1\textwidth]{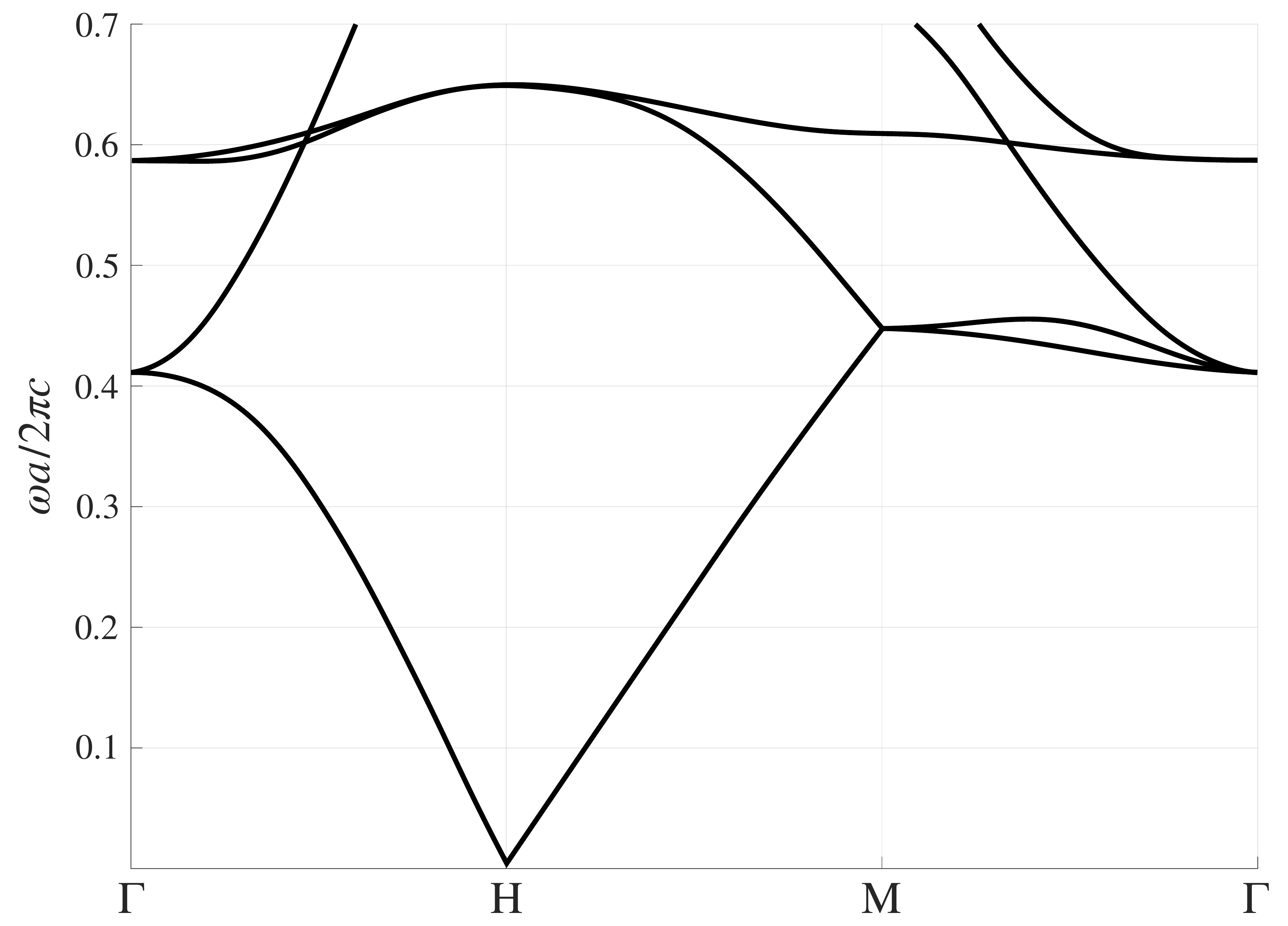}}
	\end{minipage}
	\caption{Dispersion diagram for the $\Gamma\mathrm{H}\mathrm{M}\Gamma$-path. Brillouin zone points coordinates: $\Gamma=\left(0,0,0\right)^\mathrm{T}$,  $\mathrm{H}=\left(2\pi/a,0,0\right)^\mathrm{T}$, $\mathrm{M}=\left(\pi/a,\pi/a,0\right)^\mathrm{T}$.}
	\label{fig:disp-diag}
\end{figure}

\begin{figure}[h] %--------------------------------FIGURE!
	\begin{minipage}[h]{1\linewidth}
		\center{\includegraphics[width=1\textwidth]{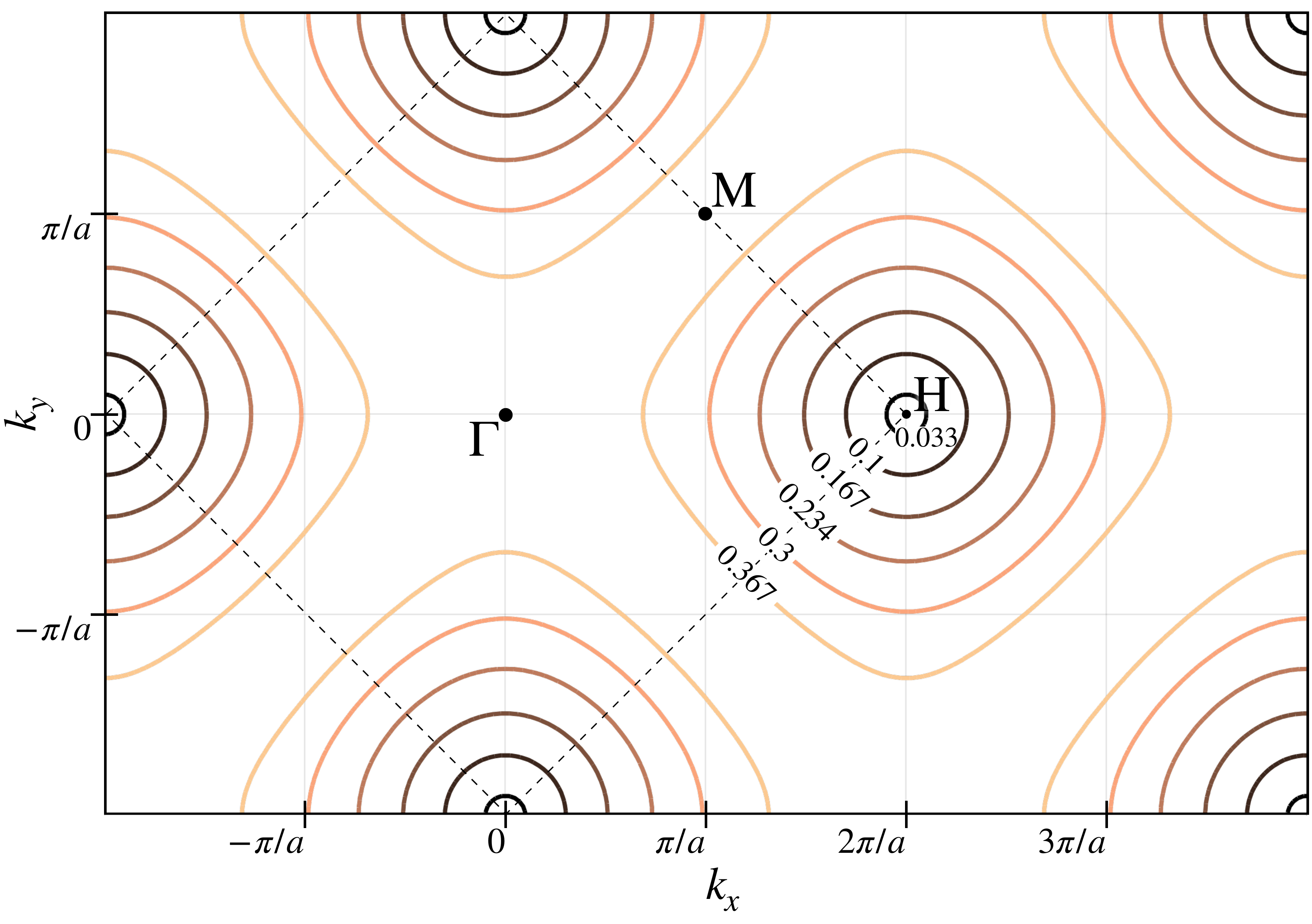}}
	\end{minipage}
	\caption{Isofrequency contours in the $k_xk_y$ plane. The frequencies on the contours are pointed in normalized units of $2\pi c/a$.}
	\label{fig:isofreq}
\end{figure}

\begin{figure}[h]%--------------------------------FIGURE!
	\begin{minipage}[h]{0.7\linewidth}
		\center{\includegraphics[width=1\textwidth]{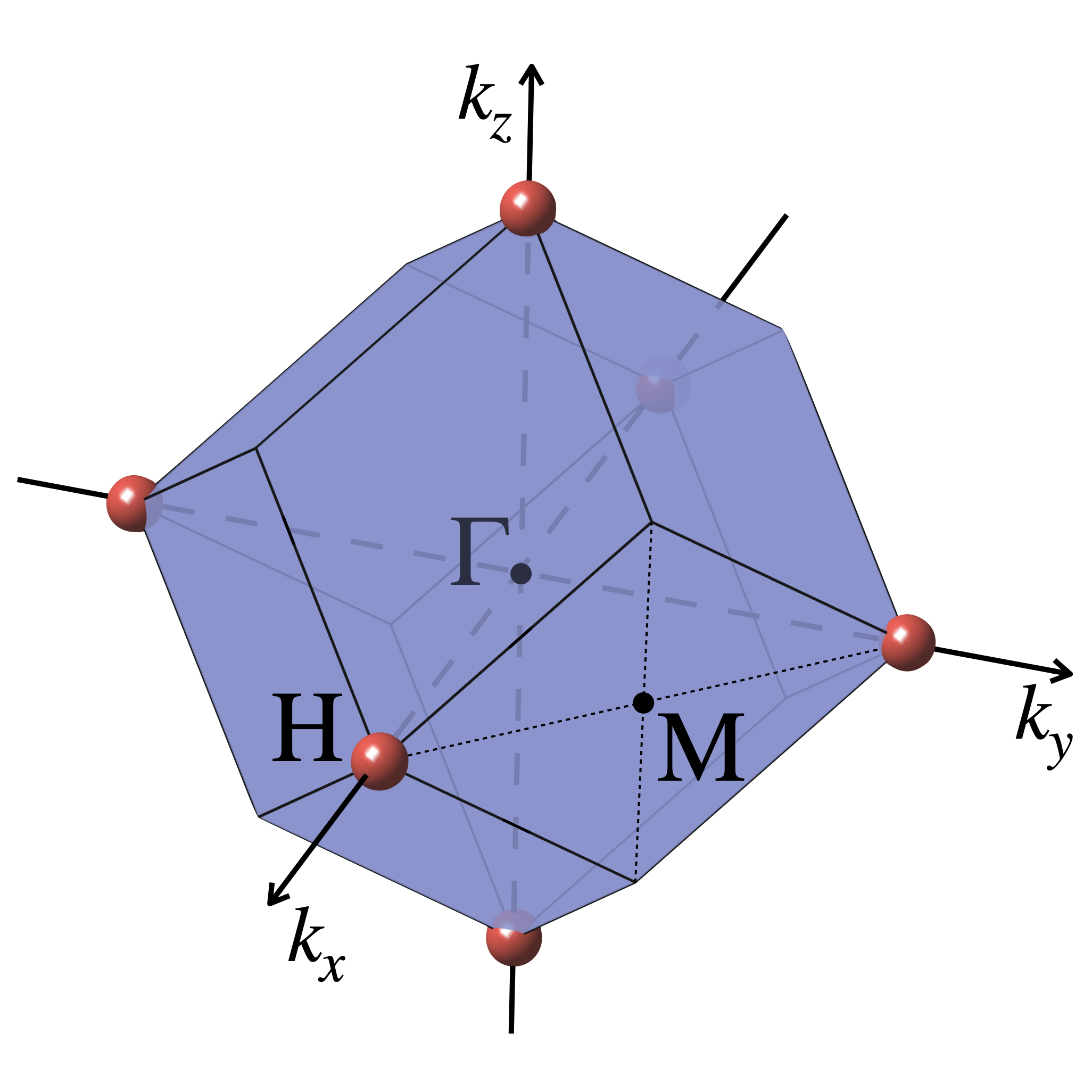} }
	\end{minipage}
	\caption{Isofrequency surfaces at the first Brillouin zone for $\omega=0.05\cdot 2\pi c/a$ (red spheres).}
	\label{fig:b-zone-spheres}
\end{figure}

We have calculated dispersion properties for the interlaced wire medium using commercial software package COMSOL Multiphysics by applying  the periodic boundary conditions (with $e^{+i\left(\vec k\cdot\vec r\right)}$ spatial dependence) to the unit cell with the wave vector $\vec k$ spanning the first Brillouin zone. As a result, we have calculated the respective eigenfrequencies $\omega(\vec k)$. The dispersion properties of the studied metamaterial are illustrated by (1) the dispersion diagram shown in  Fig. \ref{fig:disp-diag};
(2) the isofrequency contours in $k_xk_y$ plane for set of 5 frequencies shown in Fig. \ref{fig:isofreq} and (3) the isofrequency surfaces for a single frequency  shown in Fig.~\ref{fig:b-zone-spheres}.

% In $k_xk_y$ plane which is parallel to wires the coordinates of key points of the Brillouin zone in the plane are:
%\begin{equation}
%    \begin{array}{ll}
%    \Gamma=\pi/a\left(0,0,0\right)^\mathrm{T},&
%    \mathrm{X}=\pi/a\left(1,0,0\right)^\mathrm{T},\\
%    \mathrm{H}=\pi/a\left(2,0,0\right)^\mathrm{T},&
%    \mathrm{M}=\pi/a\left(1,1,0\right)^\mathrm{T}.
%    \end{array}
%    \label{eq:b-zone-vertex}
%\end{equation}

According to these results, it is discovered that at low frequencies the isofrequency surfaces of the metamaterial  surround the corners ($\mathrm{H}$ points) of the Brillouin zone, but not the $\Gamma$ point as in \cite{fan2007nonmax,silveirinha2017lighttun,Powell2021}. This means that the metamaterial supports waves with extremely large wave vectors at low frequencies. 
As one can see from dispersion diagram in Fig. \ref{fig:disp-diag} the metamaterial has artificial plasma frequency \cite{Pendry1,Pendry2,silveirinha2009plasma} around $\omega a/(2\pi c)=0.41$. However, in contrary to the case of three-dimensional connected wire medium, the bulk plasmon mode branch (which usually exists below the plasma frequency) bends down to zero at $\mathrm H$ point. Despite the unexpectedness of such results due to the seemingly isotropic nature of the material, this conclusion can be justified theoretically.

To this end, following the approach of Ref.~\cite{chen2018kpoints}, we consider the low-frequency limit when the wave equation is converted to the Poisson's equation, the latter can be written in terms of electrostatic potentials. Since two wire sublattices are not connected, they should have distinct potentials $\varphi_1$ and $\varphi_2$.

On the other hand, periodic nature of the system requires that the potential difference between the two sublattices should satisfy Bloch's theorem, which yields:
\begin{equation}
	\begin{cases}
		(\varphi_1-\varphi_2)=(\varphi_2-\varphi_1) e^{i(\mathbf{k} \cdot\boldsymbol{a}_1)}\\
		(\varphi_1-\varphi_2)=(\varphi_2-\varphi_1) e^{i(\mathbf{k} \cdot\boldsymbol{a}_2)}\\
		(\varphi_1-\varphi_2)=(\varphi_2-\varphi_1) e^{i(\mathbf{k} \cdot\boldsymbol{a}_3)}
	\end{cases}
	\label{eq:romb-system}
\end{equation}

Each of equations in (\ref{eq:romb-system}) relates the potential difference between meshes at the opposite surfaces of the rhombohedral cell. For example, Fig. \hyperref[fig:unitcell]{\ref{fig:unitcell}(b)} suggests that $A_0B_2A_1B_3$ coincides with $B_1A_3B_0A_2$ when shifted by a vector $\boldsymbol{a}_1$. Similarly, the Bloch's theorem can be applied for the other two pairs of the opposite faces.

The system of equations \hyperref[eq:romb-system]{(\ref{eq:romb-system})} is equivalent to the following conditions:
\begin{equation}
	T\cdot\mathbf{k} = 
	\begin{pmatrix}
		2n_1+1\\
		2n_2+1\\
		2n_3+1
	\end{pmatrix}\pi,\quad n_1,n_2,n_3\in\mathbb{Z}
	\label{eq:romb-system-k-1}
\end{equation} 

where 
\begin{equation}
	T = 
	\begin{pmatrix}
		\boldsymbol{a}_1^\mathrm{T}\\
		\boldsymbol{a}_2^\mathrm{T}\\
		\boldsymbol{a}_3^\mathrm{T}
	\end{pmatrix} =	a/2
	\begin{pmatrix}
		-1& 1& 1\\
		-1& 1& -1\\
		-1& -1&1
	\end{pmatrix}. 
	\label{eq:romb-translation}
\end{equation} 

Hence, it is straightforward to show that:
\begin{equation}
	\begin{pmatrix}
		k_x\\
		k_y\\
		k_z
	\end{pmatrix}=
	\begin{pmatrix}
		n_2+n_3+1\\
		n_3-n_1\\
		n_2-n_1
	\end{pmatrix}\frac{2\pi}{a},\quad n_1,n_2,n_3\in\mathbb{Z}
	\label{eq:romb-system-k-3}
\end{equation}

Depicting the set of solutions corresponding to Eq. \hyperref[eq:romb-system-k-3]{(\ref{eq:romb-system-k-3})} in reciprocal space, we observe that the isofrequency surfaces  emerge from  $\mathrm{H}$-points as in Fig. \ref{fig:b-zone-spheres}. Importantly, no isofrequency surfaces appear at the point $\Gamma$  since $k_x=k_y=k_z=0$ is not the solution of Eq.~\hyperref[eq:romb-system-k-3]{(\ref{eq:romb-system-k-3})}.

%\section{\label{sec:polariz} Polarization}

%The last important issue that will be considered in this article is the issue of polarization of the first low-frequency mode, which we have talked about up to now. In one of the early works\cite{silveirinha2017lighttun}, the longitudinal polarization of the first mode was claimed, but so far no detailed analysis has been presented to confirm this fact.

Thus, we have proved both numerically and analytically that the interlaced wire medium supports the modes with large wave vector at low frequencies. In order to determine the polarization state of of these modes, we have calculated the average electric field numerically taking into account phase shift due to wave vector~\cite{silveirinha2005homo}:
\begin{equation}
	\mathbf{E}^{av}= \frac{1}{V_{\rm unit\ cell}}\int\limits_{\rm unit\ cell} \mathbf{E}(\mathbf{r}) \cdot e^{-i \mathbf{k} \cdot \mathbf{r}} dV,
	\label{eq:av-field}
\end{equation} 
where $\mathbf{r}$ provides the coordinates of the point inside the unit cell, $\mathbf{E}(\mathbf{r})$ is the electric field at this point and $\mathbf{k}$ is the wave vector for which the eigenmode is calculated.

\begin{figure}[h]
	\begin{minipage}[h]{1\linewidth}
		\center{\includegraphics[width=1\textwidth]{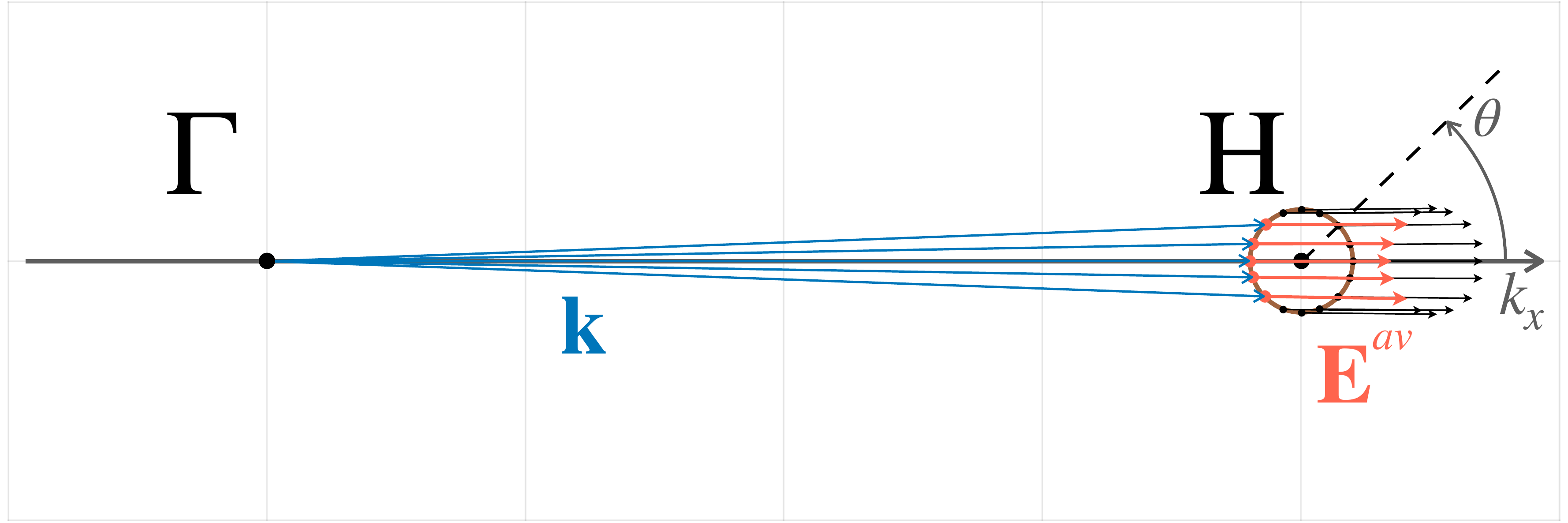} \\ (a)}
	\end{minipage}
	\vfill
		\begin{minipage}[h]{1\linewidth}
		\center{\includegraphics[width=1\textwidth]{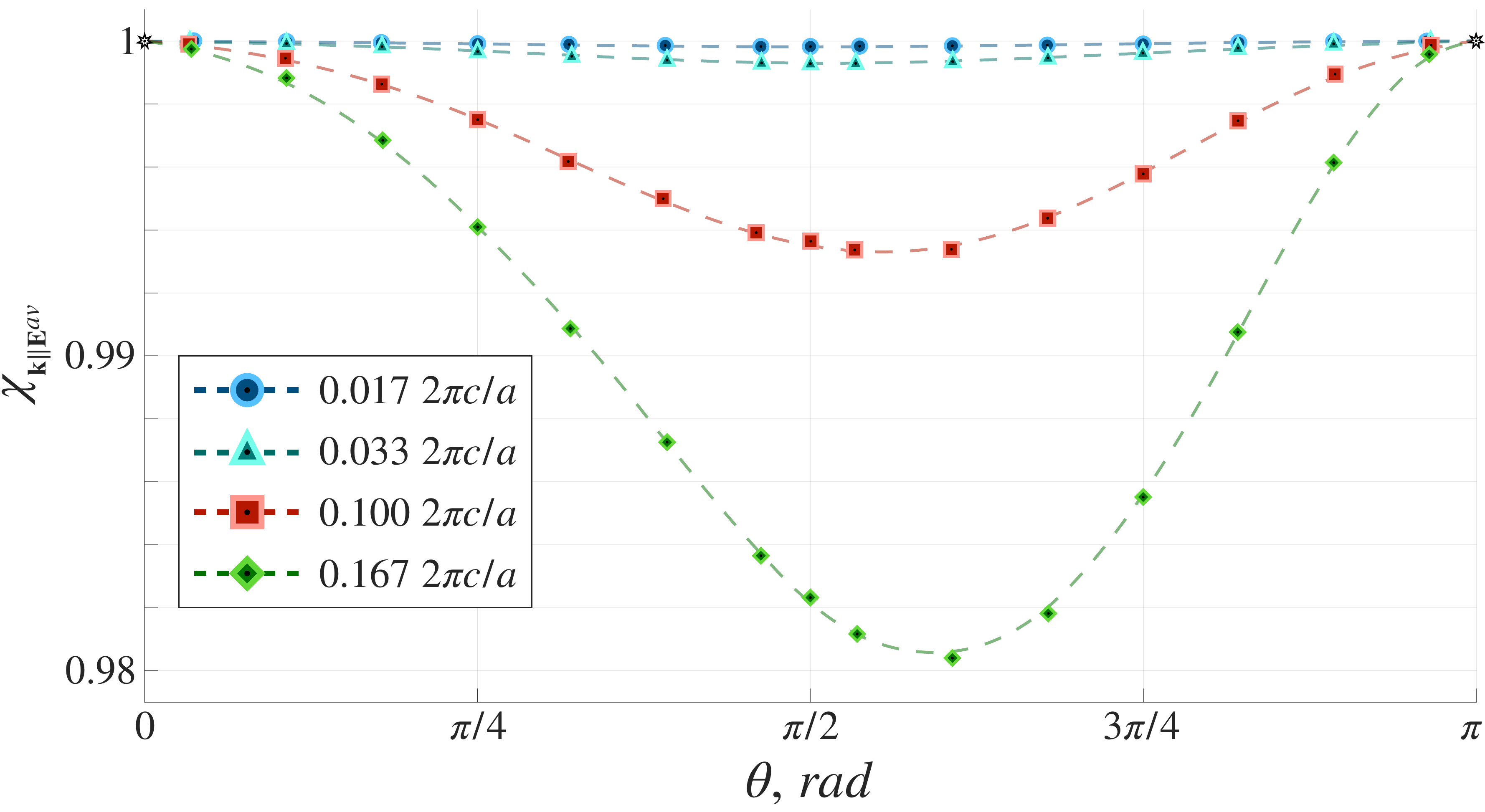} \\ (b)}
	\end{minipage}
	\caption{(a) Isofrequency contour $\omega=0.033\cdot 2\pi c/a$ with the designation of $\Gamma$ and $\mathrm{H}$ points on it and vectors of the average electric field $\vec E^{av}$ and $\vec k$. (b) Plot of the dependence of the longitudinal wave ratio $\chi_{{}_{\mathbf{k}\parallel \mathbf{E}^{av}}}$ on the angular coordinate $\theta$ of a point on the contour (the angle between $k_x$ and the vector drawn from $\mathrm{H}$-point) for different frequencies.}
	\label{fig:polarization}
\end{figure}

The result of averaging of the electric field for the wave vectors corresponding to a single low-frequency isofrequency contour is illustrated in Fig. \hyperref[fig:unitcell]{\ref{fig:polarization}(a)}.
One can clearly see that the mode is longitudinal, i.e. the vector $\mathbf{k}$ is parallel to the vector $\mathbf{E}^{av}$.  In order to accurately assess the polarization of the mode we calculated the longitudinal coefficient:
\begin{equation}
	\chi_{{}_{\mathbf{k}\parallel \mathbf{E}^{av}}} = \frac{\left(\mathbf{k}\cdot \mathbf{E}^{av}\right)}{\left| \mathbf{k}\right| \left|\mathbf{E}^{av}\right|}
	\label{eq:long-wave}
\end{equation} 
as a function of the angular coordinate $\theta$ of a point on the contour at a given frequency, see Fig. \hyperref[fig:unitcell]{\ref{fig:polarization}(b)}.
By the definition, the closer the longitudinal coefficient to one, the smaller the angle between the electric field and the wave vector.

As one can see from Fig. \hyperref[fig:unitcell]{\ref{fig:polarization}(b)} with the decrease of the frequency the curve of longitudinal coefficient approaches to 1. The maximum deviation from the longitudinal polarization is less than 2\% for  $\omega=0.167\cdot 2\pi c/a$ and less than 1\% for $\omega=0.1\cdot 2\pi c/a$. Thus, low-frequency modes of interlaced wire medium are longitudinal.

\begin{figure}[h] %--------------------------------FIGURE!
    \begin{minipage}[h]{0.493\linewidth}
		\center{\includegraphics[width=1\textwidth]{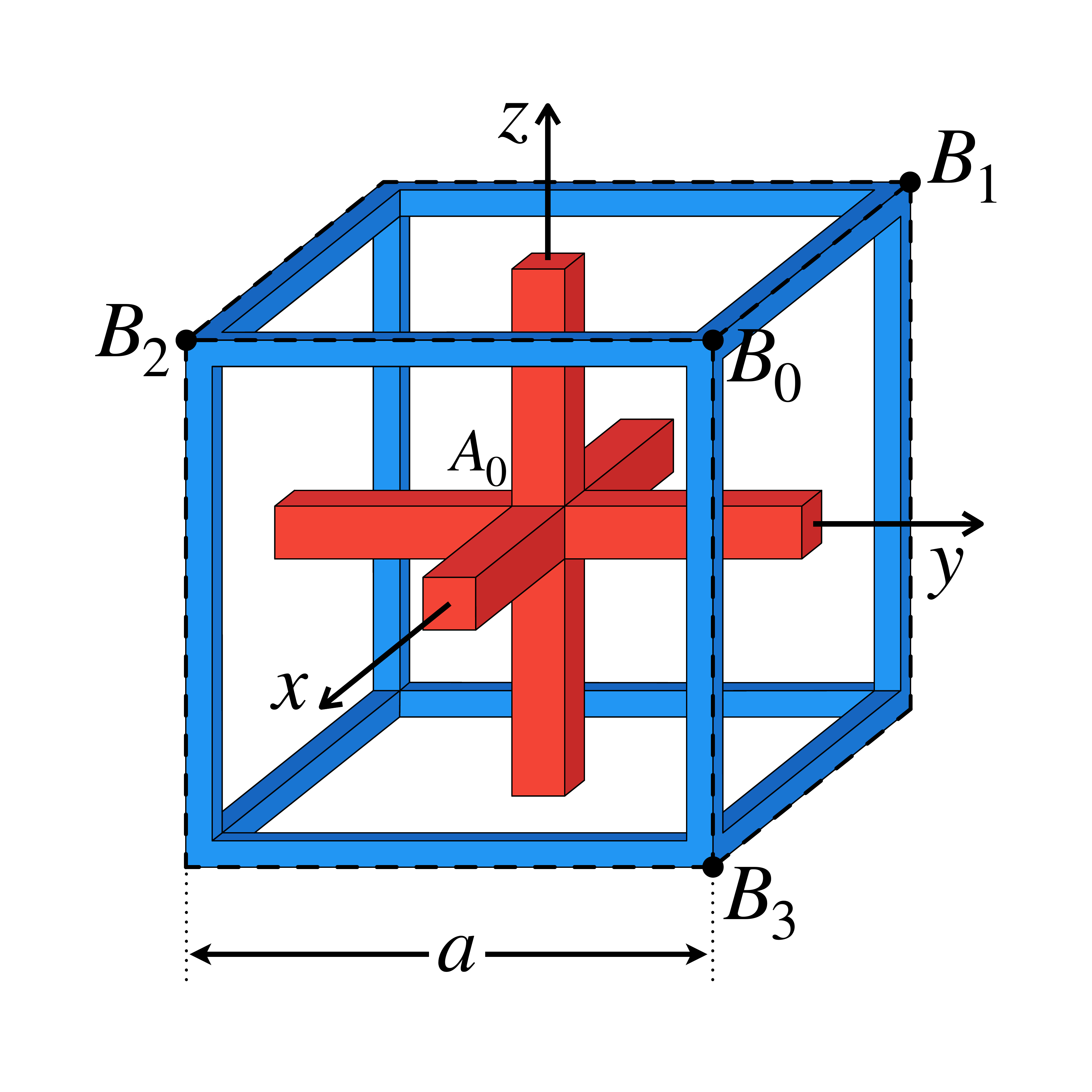}\\ (a)}
	\end{minipage}
	\hfill
	\begin{minipage}[h]{0.493\linewidth}
		\center{\includegraphics[width=1\textwidth]{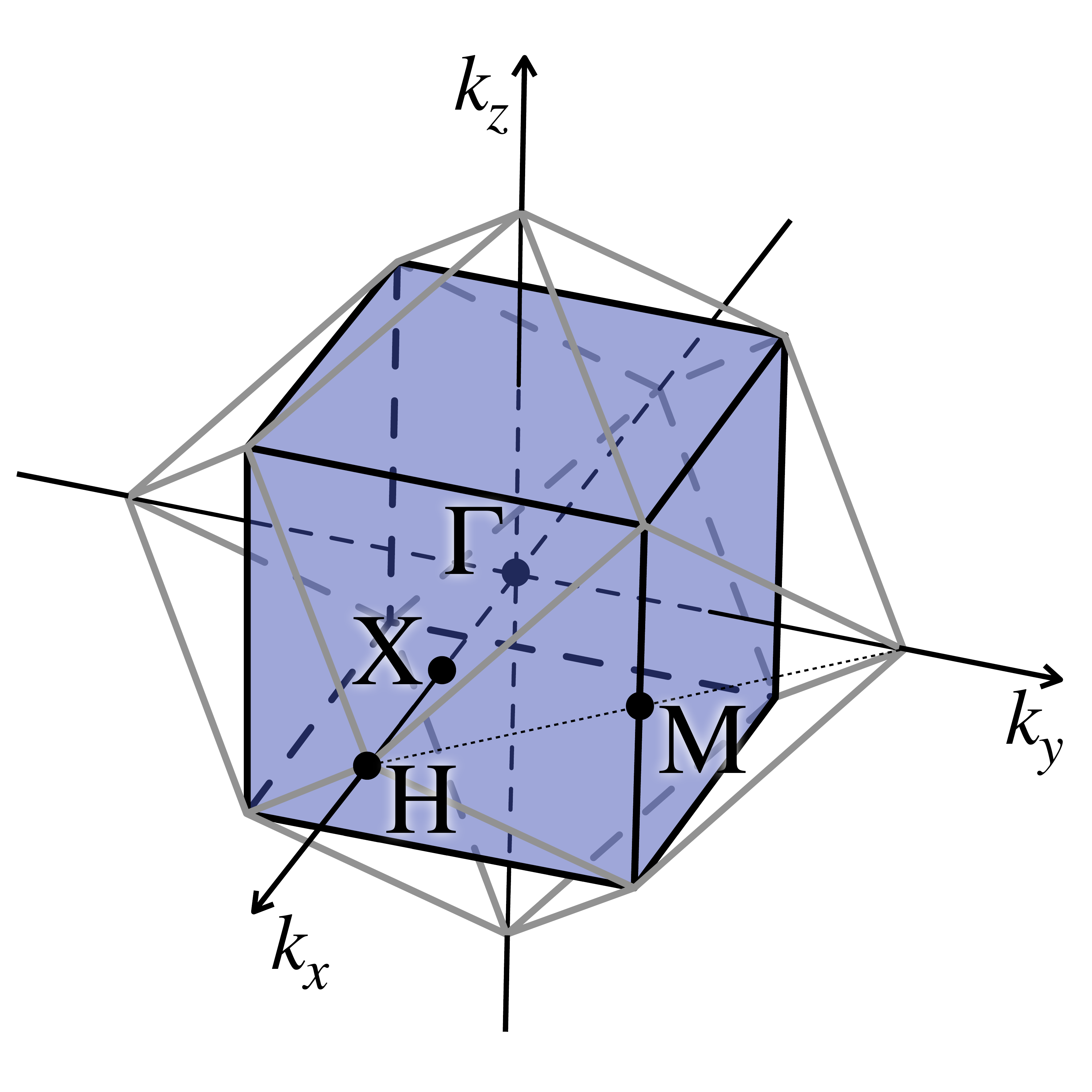}\\ (b)}
	\end{minipage}
	\caption{(a) Cubic supercell of the metamaterial. (b) Cubic Brillouin zone corresponding to the supercell inside the rhombic dodecahedron Brillouin zone for the primitive unit cell. Points coordinates: $\Gamma=\left(0,0,0\right)^\mathrm{T}$, $\mathrm{X}=\left(\pi/a,0,0\right)^\mathrm{T}$, $\mathrm{H}=\left(2\pi/a,0,0\right)^\mathrm{T}$, $\mathrm{M}=\left(\pi/a,\pi/a,0\right)^\mathrm{T}$.}
	\label{fig:supercell}
\end{figure}

\begin{figure*}[t] %--------------------------------FIGURE!
	\begin{minipage}[h]{0.8\linewidth}
		\center{\includegraphics[width=1\textwidth]{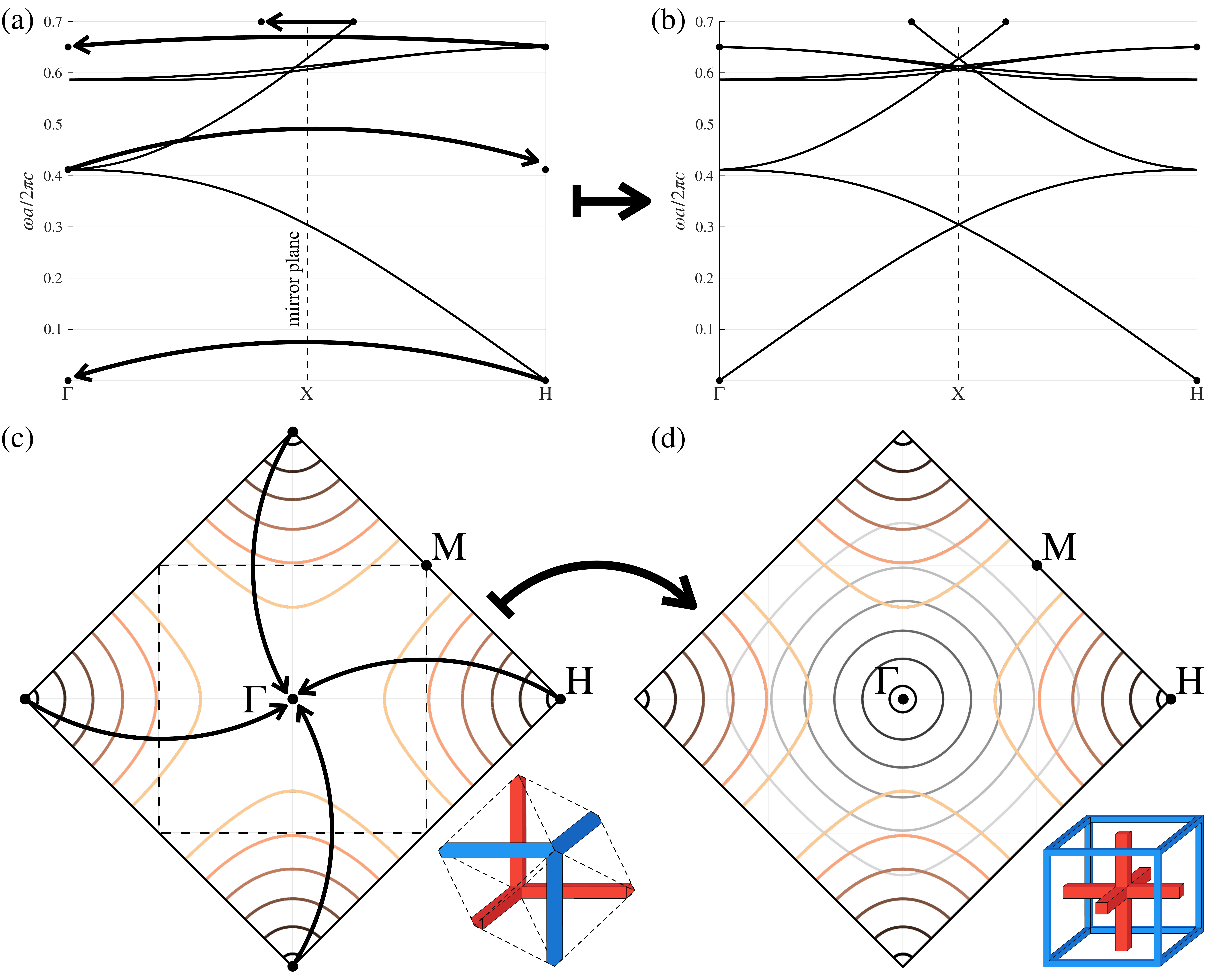}}
	\end{minipage}
	\caption{(a) Dispersion diagram for the $\Gamma\mathrm{X}\mathrm{H}$-path (same as in Fig. \ref{fig:disp-diag}) for the primitive unit cell  and the flipping scheme for obtaining (b) the dispersion diagram for the cubic super cell. (c) Isofrequency contours calculated for the primitive unit cell (same as in Fig. \ref{fig:isofreq}) and the flipping (origami folding) scheme for obtaining (d) the isofrequency contours for the cubic super cell. $\Gamma=\left(0,0,0\right)^\mathrm{T}$, $\mathrm{X}=\left(\pi/a,0,0\right)^\mathrm{T}$, $\mathrm{H}=\left(2\pi/a,0,0\right)^\mathrm{T}$, $\mathrm{M}=\left(\pi/a,\pi/a,0\right)^\mathrm{T}$.}
	\label{fig:flipping}
\end{figure*}

In the papers \cite{fan2007nonmax,Powell2021} the authors considered the same interlaced wire medium structure as in this Letter, but it was assumed that the unit cell of the structure has cubic shape as in Fig. \hyperref[fig:supercell]{\ref{fig:supercell}(a)}. The cubic super cell is two times larger in volume than the primitive unit cell (Fig. \hyperref[fig:unitcell]{\ref{fig:unitcell}(b)}). This doubling of primitive unit cell results in 
double decrease of volume of Brillouin zone (Fig. \hyperref[fig:unitcell]{\ref{fig:unitcell}(c)}) and change from rhombic dodecahedron shape to cubic shape as demonstrated in Fig. \hyperref[fig:flipping]{\ref{fig:flipping}(b)}. 
The dispersion diagram along $\Gamma \mathrm{X}$ path for the double supercell can be obtained from corresponding dispersion diagram (Fig. \ref{fig:disp-diag}) for the primitive unit cell by operation of mirror reflection (flipping) as shown in Fig. \hyperref[fig:flipping]{\ref{fig:flipping}(a,b)}. 
The flipping of the dispersion diagram is described in Ref. \cite{kaina2015negative} where doubling of unit cell is studied in details.
Note, that this dispersion diagram is the same as ones presented in Fig. 1(b) of Ref. \cite{fan2007nonmax} and in Fig. 1(f) of Ref. \cite{Powell2021}, and the dispersion diagram contains a brunch starting from $\Gamma$ point at low frequencies.
The isofrequency contours for the double supercell can be obtained from corresponding isofrequency contours (Fig. \ref{fig:isofreq}) for the primitive unit cell by operation of origami folding (flipping) as shown in Fig. \hyperref[fig:flipping]{\ref{fig:flipping}(c,d)}.

It is simpler (both numerically and analytically) to consider interlaced wire medium as material with cubic unit cell as in Fig. \hyperref[fig:supercell]{\ref{fig:supercell}(a)} and Brillouin zone as in Fig. \hyperref[fig:supercell]{\ref{fig:supercell}(b)}, but
this approach does not allow to describe dispersion properties of the interlaced wire metamaterial completely.
For example, the use of double supercell does not allow
to distinguish such points in the reciprocal space as $\Gamma$ and $\mathrm{H}$.
Due to Bloch theorem these points are equivalent within double supercell description, but they are actually separate points of the Brilloun zone corresponding to the primitive unit cell (Fig. \hyperref[fig:unitcell]{\ref{fig:unitcell}(c)}). 
However, as it was shown above, the metamaterial supports modes with large wave vectors (close to $\mathrm{H}$ point) but does not support modes with small wave vectors (close to $\Gamma$ point) as it may seem from dispersion diagrams (Fig. \hyperref[fig:flipping]{\ref{fig:flipping}}(b)) and isofrequency contours (Fig. \hyperref[fig:flipping]{\ref{fig:flipping}}(d)) for cubic supercell.

In order to illustrate importance of use primitive unit cell instead of cubic supercell we have analysed spatial spectrum of typical eigenmodes supported by the metamatertial at low frequencies. The amplitudes of Bloch harmonics $\vec E_{m,n,l}$ of an eigenmode $\vec E(\vec r)$ were numerically calculated :  
\begin{equation}
	\mathbf{E}_{m,n,l}= \frac{1}{V_{\rm super\ cell}}\int\limits_{\rm super\ cell} \mathbf{E}(\mathbf{r}) \cdot e^{-i \mathbf{k_{m,n,l}} \cdot \mathbf{r}} dV,
	\label{eq:harmonic}
\end{equation}
where $\vec k_{m,n,l}=\vec k + 2\pi/a (m,n,l)^T$, $m,n,l \in \mathbb{Z}$.

\begin{figure}[h!] %--------------------------------FIGURE!
	\begin{minipage}[h]{0.8\linewidth}
		\center{\includegraphics[width=1\textwidth]{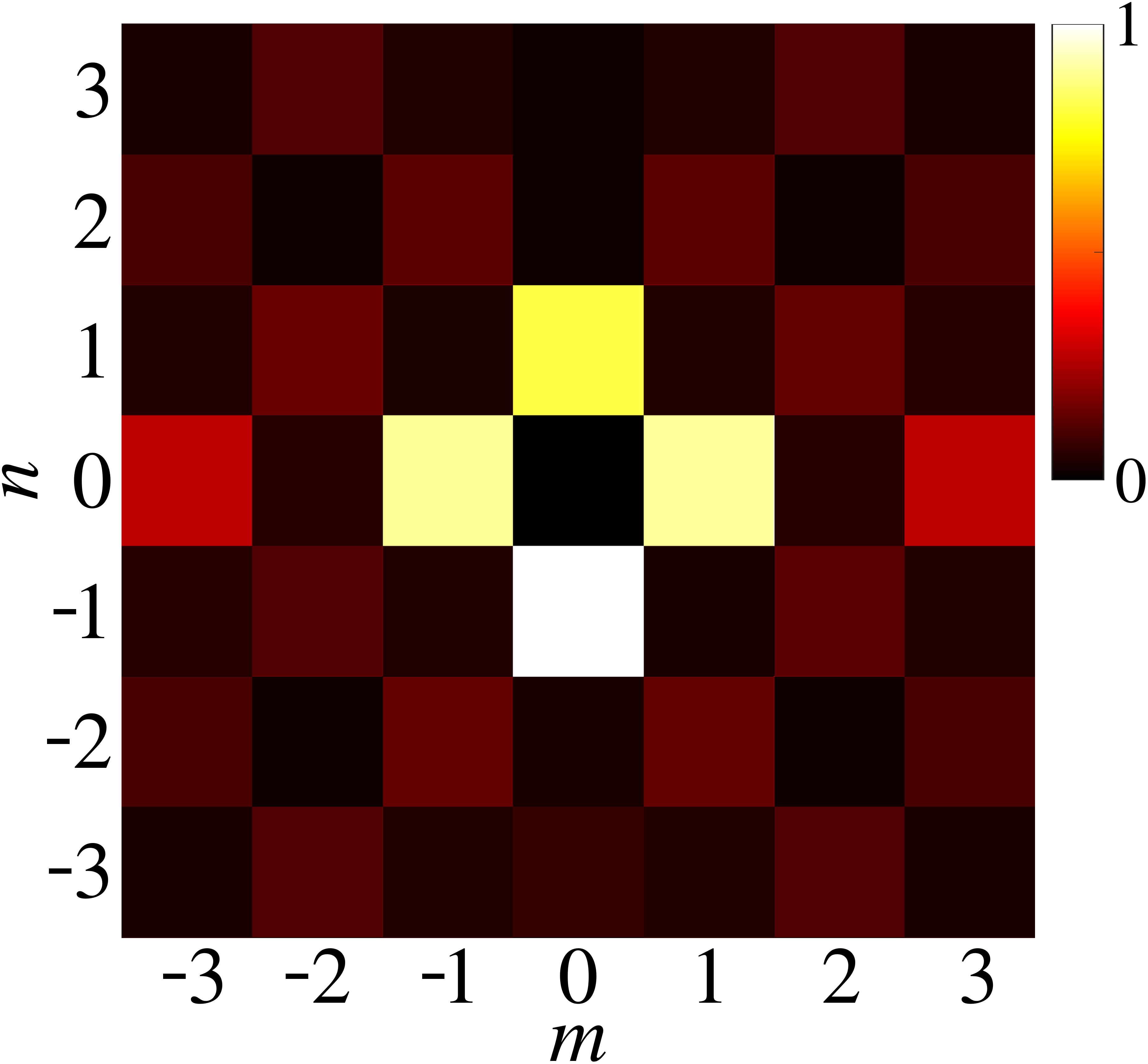}}
	\end{minipage}
	\caption{Histogram of Bloch harmonics amplitudes $|\vec E_{m,n,0}|$ (normalized by maximum value) for an eigenmode corresponding to $\omega a/2\pi c=0.1$ and $\vec k=(0, 0.296, 0)^T\pi/a$.}
	\label{fig:pwe-diag}
\end{figure}

The spatial spectrum $\vec E_{m,n,l}$ of an eigenmode with $\omega a/2\pi c=0.1$ and $\vec k=(0, 0.296, 0)^T\pi/a$ is shown in Fig. \ref{fig:pwe-diag} for $m=-3...3$, $n=-3...3$, $l=0$. One can see that the eignemode has large amplitudes $\vec E_{0,-1,0}, \vec E_{-1,0,0}, \vec E_{0,1,0}$ and $\vec E_{1,0,0}$ while $\vec E_{0,0,0}$ is negligibly small. 
This means that the mode of the metamaterial does not have a spatial harmonic with wave vector close to $\Gamma$ point in its spatial spectrum. The spectrum is dominated by harmonics with large wave vectors close to points described by Eq. (\ref{eq:romb-system-k-3}). Our numerical calculations demonstrate that such behavior of spatial spectrum is typical for eignemodes of interlaced wire metamaterial at low frequencies and does not significantly depend on direction of wave vector $\vec k$. 

%This means that the eigenmode has nearly zero average field $\vec E^{av}(\omega, (0, 0.1, 0)^T)=\vec E_{0,0,0}$ within supercell while it has $\vec E^{av}(\omega, \vec (0, 0.1-2\pi/a, 0)^T)=\vec E_{0,-1,0}$ within primitive unit cell.

In conclusion, in this Letter we have studied the dispersion properties of interlaced wire medium in the symmetric configuration. As we have proved, this metamaterial supports low-frequency modes with large wave vectors and longitudinal polarization, which highlights strongly nonlocal response of the structure. We believe that the control on spatial dispersion effects in metamaterials will enable promising applications such as recently demonstrated all-angle impedance matching~\cite{ImpedanceMatch2018,Nonlocality2018}, polarization control \cite{Powell2021}, imaging with subwavelength resolution \cite{WMlens} or squeezing the wavelength of electromagnetic fields, as suggested in this Letter, to enable forbidden transitions~\cite{Rivera2016}.

\section{Acknowledgement}
The authors are grateful to Maxim Gorlach for very fruitful discussion.

\bibliographystyle{ieeetr}
\bibliography{bib}
\end{document}